\documentclass[%
preprint,
 amsmath,amssymb,
 aps,
prl,
floatfix,
]{revtex4-1}

\usepackage{graphicx}% Include figure files
\usepackage{dcolumn}% Align table columns on decimal point
\usepackage{bm}% bold math
\usepackage{subfigure}
\begin{document}

\preprint{APS/123-QED}

\title{Temperature Dependent Layer Breathing Modes in Two Dimensional Materials}% Force line breaks with \\

\author{Indrajit Maity}
\author{Prabal K Maiti}%
%\altaffiliation[Also at]{%
% Authors' institution and/or address}%
\author{Manish Jain}
\email{mjain@iisc.ac.in}
\affiliation{
 Department of Physics, Indian Institute of Science, Bangalore-560012
}%

\date{\today}% It is always \today, today,

\begin{abstract}
Relative out of plane displacements of the constituent layers of two
dimensional materials gives rise to unique low frequency breathing modes. By
computing the height-height correlation functions in momentum space, we show
that, the layer breathing modes (LBMs) can be mapped consistently to vibrations
of a simple linear chain model. Our calculated thickness dependence of LBM
frequencies for few layer (FL) graphene and molybdenum disulphide (MoS$_{2}$)
are in excellent agreement with available experiments. Our results show a
redshift of LBM frequency with increase in temperature, which is a direct
consequence of anharmonicities present in the interlayer interaction. We also
predict the  thickness and temperature dependence of LBM frequencies for FL
hexagonal boron nitride (hBN). Our study provides a simple and efficient way to
probe the interlayer interaction for layered materials and their
heterostructures, with the inclusion of anharmonic effects.      
\end{abstract}
% \begin{description}
% Secondary publications and information retrieval purposes.
% \item[PACS numbers]
% May be entered using the \verb+\pacs{#1}+ command.
% You may use the \texttt{description} environment to structure your abstract;
% use the optional argument of the \verb+\item+ command to give the category of each item. 
%\end{description}

\pacs{Valid PACS appear here}% PACS, the Physics and Astronomy
\maketitle

%\tableofcontents

%\section{\label{sec:level1}First-level heading}
Two dimensional (2D) materials, for example, graphene, transition metal
dichalcogenides, hBN, are being studied extensively for their exciting
electronic, thermal, mechanical properties \cite{novoselovroadmap,
beyondgraphene}. A great deal of effort has also been directed towards
understanding hybrid structures of these 2D materials \cite{geimhetero}. It is
well known that, typically few layers of 2D materials and their hybrid
structures are coupled by weak van der Waals (VDW) forces. Such layer-layer
couplings give rise to unique low frequency \textit{interlayer} vibrational
modes at finite temperature, namely, shear and layer breathing modes (LBMs).
\cite{tanshear,review_acsnano}. It has been found experimentally
that, LBMs are more sensitive to external perturbations than shear modes
\cite{luitemperature}. These LBMs can be used as direct probe to determine
layer thickness, stacking order, effects of external  environment, adsorbates
etc
\cite{luiLBM1,luitemperature,hegraphene,zhangMoS2,zhaoMoS2,boukhichaMoS2,yanMoS2,lingP,zhaoBi2Te3,mos2_graphene,heNbSe2,heReS2,luivdw}.
Furthermore, LBMs play a crucial role in interlayer electric conductance \cite{conductance}, thermoelectric transport \cite{phani_nano}. Understanding the origin and quantification of LBM
frequencies is thus of immense practical importance. 

Three key features emerge from the low frequency Raman spectroscopic
measurements of LBMs in 2D materials : (i) A system with $n$ layers will have
$n-1$ \textit{distinct} LBMs \footnote{ There is no periodicity in the out of
plane (z) direction of FL system. $\Gamma - A$ branch of the bulk counterpart
is non existent in the Brillouin zone (BZ). One can show, the frequencies of
LBMs at $\Gamma$ point for FL system, are associated with vibrations of their
bulk correspondent along $\Gamma - A$ direction}. (ii) LBM frequencies (at the
$\Gamma$ point) are highly sensitive to the thickness of the material i.e.
number of layers. For instance, when the number of layers of graphene is
increased from 2 to 8, the lowest LBM frequency redshifts from 81 cm$^{-1}$ to
22 cm$^{-1}$ \cite{luitemperature}. (iii) The lowest LBM frequency also
redshifts with increment of temperature ($T$), as seen in experiments by
controlled laser heating \cite{luitemperature,lingP}. The reported linewidths
in Raman spectroscopic measurements for LBMs are typically larger than shear
modes \cite{boukhichaMoS2}. These observations suggest the presence of strong
anharmonicity in the interlayer interaction for LBMs. In this work, we address
these three key aspects of LBMs.   

A 2D material embedded in 3D space can have out of plane acoustic phonon modes called
flexural modes (ZA). In the harmonic approximation, these flexural modes have a dispersion,
$\omega_{flex} \propto q^{2}$ for small momentum, $q$. 
For $n$ layers, due to interlayer coupling,
the degeneracy in ZA branch is lifted and  \textit{distinct} modes appear in
the vibrational spectra, implying vertical stretching/compression of the
layers. These modes are known as LBMs (ZO$^{\ensuremath{'}}$, optical modes).
In order to understand the thickness dependence of LBMs, two common approaches
are used. First, a linear chain model of $n$ masses with nearest neighbor
interaction is used widely to determine LBM frequencies. This simple model has
been shown to predict the frequencies accurately, given a knowledge of nearest
neighbor layer coupling \cite{luitemperature, zhaoMoS2, luiLBM1, zhaoMoS2, boukhichaMoS2}. However, the
mapping of the $n$ layer system to such a simple model starting from a more
general description of the constituent layers is unclear. The effects of
next-nearest neighbor layer coupling in such a model have not been quantified as well.
Second, first principles calculations based on density functional perturbation
theory (DFPT) are frequently used to calculate LBM frequencies
\cite{sahaphonons,zhaoMoS2}. In these calculations, however, the temperature
dependence of LBMs is not revealed. The inclusion of anharmonic effects i.e.
multi-phonon processes and thermal expansion coefficients are necessary to
capture the temperature dependence of LBM frequencies.  

Here, we present a simple method to calculate LBM frequencies by using a
combination of classical molecular dynamics (MD) simulations and theory of membranes. We
justify the application of linear chain model in the small momentum regime ($q\to 0$), by computing the height-height correlations. Our calculations of
layer dependence of LBM frequencies for few-layer graphene and MoS$_2$ are in
excellent agreement with available experiments. We show the evolution of LBM
frequency with temperature  for bilayer (BL) system of graphene, MoS$_{2}$ and
hBN. In the studied temperature ($T$) range, we find expansion of interlayer
separation and redshift in LBM frequency with $T$ increment. As, the interlayer
separation is calculated directly from MD simulation, \textit{all}
anharmonicities in the interlayer interaction are incorporated in the calculation.

We perform MD simulations with periodic boundary condition in the $NPT$ ensemble using
Nos\'e-Hoover thermostat and barostat as implemented in LAMMPS \cite{plimpton}.
We simulate three different layered materials, namely, graphene, MoS$_2$, h-BN
and vary the number of layers from 2 to 6. Initially, all the samples are
chosen to be roughly square shaped and contain $\approx$ 8000-9000 atoms per
layer ($N$). After equilibration, we use 4000 snapshots (2 nanosecond
production run) to average the calculated properties. We use different
forcefields (FFs) to compute LBM frequencies. For graphene three different FFs
are adopted : Long Range Bond Order Potential for Carbon (LCBOP)
\cite{los2003}, a combination of Reactive Empirical Bond Order potential and
Lennard-Jones potential (REBO+LJ) \cite{brennersecond,graphene_lj} and
Dreiding, a more generic FF \cite{mayo}. For the case of MoS$_{2}$ and
hBN, a mix of Stillinger-Weber and Lennard-Jones potential (SW+LJ)
\cite{jiangsw,liang2009,liang2012erratum} and Dreiding are used, respectively. 

The applicability of the theory of membranes (a continuum description) to
understand long-wavelength physics in 2D materials, such as graphene, is now
well established \cite{nelsonbook,fasolinoripples,amorim}. In the harmonic
approximation of membrane theory, the bending energy for a BL system with  weak VDW interaction
between the layers, can be written  as, 
\begin{equation}
E_{BL} = \frac{1}{2}\int \big[\kappa (\nabla^{2}h_{1})^{2} + \kappa (\nabla^{2}h_{2})^{2} + \sigma (h_{1}-h_{2})^{2}\big]d^{2}x
\end{equation}
where $\kappa$ is the bending rigidity of each constituent layer, $h_{1}$,
$h_{2}$ are heights of two layers with respect to each of their reference plane
and $\sigma$ denotes the interlayer coupling. In the momentum space, using the
combinations $h=(h_{1} + h_{2})/\sqrt{2}$ and $\delta h =
(h_{1}-h_{2})/\sqrt{2}$, one can identify two modes : \textit{mean} and
\textit{fluctuation} mode. The corresponding height correlation functions
\footnote{This is the height-height correlation in momentum space. For
convenience, we write it as height correlation function.}, are 
\begin{equation}
H^{BL}(q) = \langle|h(q)|^{2}\rangle = \frac{Nk_{B}T}{S_{0}\kappa q^{4}} 
\end{equation} 
\begin{equation}
\delta H^{BL}(q) = \langle |\delta h(q)|^{2} \rangle = \frac{Nk_{B}T}{S_{0}(\kappa q^{4} + 2\sigma)}
\end{equation}
where $S_{0}$ is the surface area per atom and $q=|\vec{q} |$, is defined by the
dimension of the simulation box. The dispersion relations for the
long-wavelength physics, can be inferred from the above relations :
$\omega_{mean} = \sqrt{\frac{\kappa}{\rho}}q^{2}$ and
$\omega_{fluc}=\sqrt{\frac{\kappa q^{4} + 2\sigma}{\rho}}$, where $\rho$ is the two dimensional mass density [See Supplementary
Information (SI); Section A for single layer sheet and B for bilayer system].
It should be noted that, quantum effects are neglected in the calculation of height correlation functions
($H^{BL}(q)$, $\delta H^{BL}(q)$). While the effects are important at low $T$,
these effects are reported to be unimportant above a crossover temperature,
$T^{*} \sim$ 70-90 K \cite{amorim}. All the correlation functions presented
here, are calculated for $T \geq$ 150 K, hence, quantum effects can be
neglected.

Fig.1(a) shows height correlation functions per atom ($H^{BL}(q)/N$,
$\delta H^{BL}(q)/N$) for the \textit{mean} and \textit{fluctuation} modes in
BL graphene and MoS$_2$ at room temperature. In the figure we have shown the results for BL graphene using REBO+LJ and for BL MoS$_2$ using SW+LJ. However, the main features 
of the height correlation functions are insensitive to the choice of forcefields.
The \textit{mean} mode of BL graphene is well
described within the harmonic approximation (Eqn.(2)) for $0.5 \text{\AA}^{-1}
\leq q \leq 1.0\text{\AA}^{-1}$. The membrane theory predicts a change in
scaling, from $H^{BL}(q)\propto q^{-4}$ to $H^{BL}(q)\propto q^{-3.18}$, when
anharmonicities become important owing to the coupling of bending and
stretching \cite{leself}. This deviation from the harmonic approximations of
membrane theory i.e. a change of scaling from $H^{BL}(q)\sim q^{-4}$, is found
in all the simulated samples.  Our results show that, anharmonic effects are
more pronounced in BL MoS$_2$, compared to that of graphene (Fig.1(a)). More
generally, we find \textit{mean} mode of BL system behaves like a single layer for all the simulated materials . The
\textit{fluctuation} mode for both BL graphene and MoS$_2$ becomes a constant
for $q \lesssim 0.2 \text{\AA}^{-1} $. This implies that near the zone center
($\Gamma$ point) the interlayer coupling ($\sigma$) dictates the height
fluctuations, as predicted by Eqn.(3). This aspect of the \textit{fluctuation}
mode is key for the rest of our work. Contrary to the \textit{mean} mode, for
small $q$,  the anharmonicities arising from the coupling between bending and
stretching are found to be irrelevant for the \textit{fluctuation} mode. The \textit{fluctuation} mode is identified with LBM. The
Bragg peaks (Fig.1(a)) signify the underneath crystal lattice structure and
breakdown of membrane theory.   

For $q \to 0 $, $\omega_{mean} \to 0$ and $\omega_{fluc} \to
\sqrt{\frac{2\sigma}{\rho}}$ ; We identify $\omega_{fluc}$ as the LBM frequency
(ZO$^{\ensuremath{'}}$) of a BL system. This \textit{dispersion-less} feature
of $\omega_{fluc}$ help us in two significant ways : (i) We can estimate
$\sigma$ directly from the flat region of $\delta H^{BL}(q)$, without depending
on any other mechanical parameters. (ii) The mapping of the BL system to linear
chain model (Fig.1(b)) becomes transparent. In such a model, the force
constants are determined solely from the interlayer coupling. The schematics of
the modes of the constituent layers at the $\Gamma$ point are shown in the inset of
Fig.1(a). The interlayer interaction lifts the degeneracy of the flexural modes
of each layer into $\omega_{mean}$ and $\omega_{fluc}$ for $q \to 0$. This can
be confirmed from the differences of $\delta H^{BL}(q)/N$ and $H^{BL}(q)/N$. In
table I, we show the force constants for BL graphene and MoS$_2$ and compare those with the values obtained from first principles calculations. As can be easily examined from the table, our results are in excellent agreement with earlier
reports.  
  
\begin{figure}[htp]
    \centering
    \subfigure[]{\includegraphics[scale=0.35]{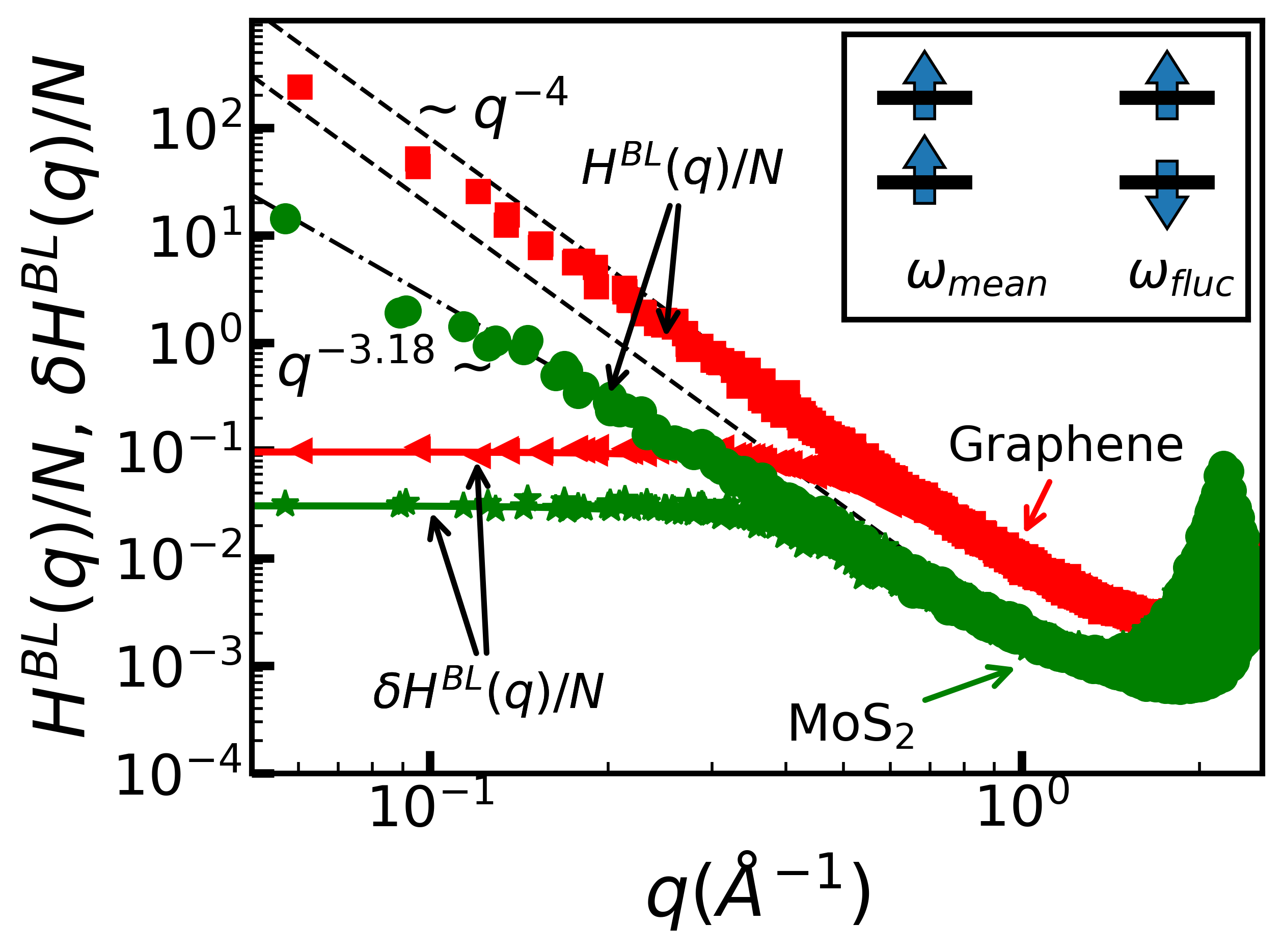}}\quad 
    \subfigure[]{\includegraphics[scale=0.17]{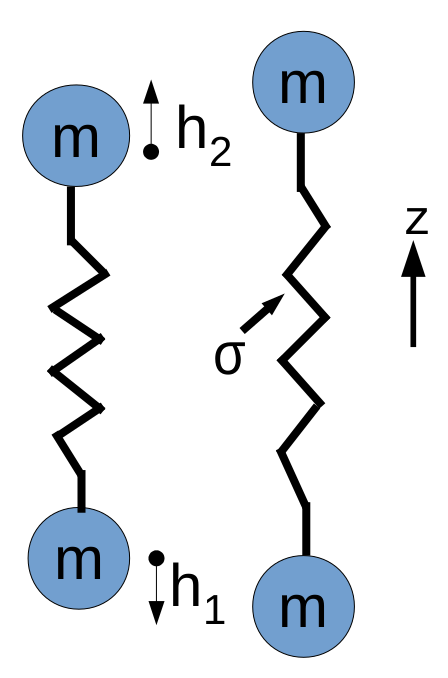}}
\caption{(a) Height correlation functions for the \textit{mean} mode,
$H^{BL}(q)/N$ (graphene : red square, MoS$_2$ : green circle) and
\textit{fluctuation} mode, $\delta H^{BL}(q)/N$ (graphene : red triangle,
MoS$_2$ : green star) for BL graphene and BL MoS$_2$. Black dashed and
dash-dotted line show scaling $q^{-4}$, $q^{-3.18}$ respectively. The solid lines denote fit to the \textit{fluctuation} mode. The inset
shows schematic of normal modes at the $\Gamma$ point. (b) The linear chain
model : two masses (m) connected by a spring with spring constant $\sigma$.}
\end{figure}

\begin{table}[!htp]
    \caption{Comparison of force constants calculated from MD simulation and first principles approach. }
    \centering
    \begin{tabular}{|*{4}{c|}}
    \hline 
     \multicolumn{1}{|c|}{BL system} &\multicolumn{1}{|c|}{Temperature} & \multicolumn{1}{|c|}{$\sigma$ (x$10^{19}$ N m$^{-3}$)} & \multicolumn{1}{|c|}{Method} \\ \hline
    Graphene & 300 K & 8.1 & REBO + LJ \\ 
       &  & 7.3 & LCBOP  \\ \hline
      Graphene & 0 K & 7.9 & DFPT \cite{sahaphonons}  \\ \hline
     MoS$_2$ & 300 K & 8.3 & SW + LJ \\ \hline
       MoS$_2$ & 0 K & 9.26 & DFPT \cite{zhaoMoS2} \\ \hline 
    \end{tabular}
\end{table}
 
\begin{figure*}[htp]
    \centering
%    \subfigure[]{\includegraphics[scale=0.285]{lowest_lbm_branch.png}}\quad 
    \subfigure[]{\includegraphics[scale=0.38]{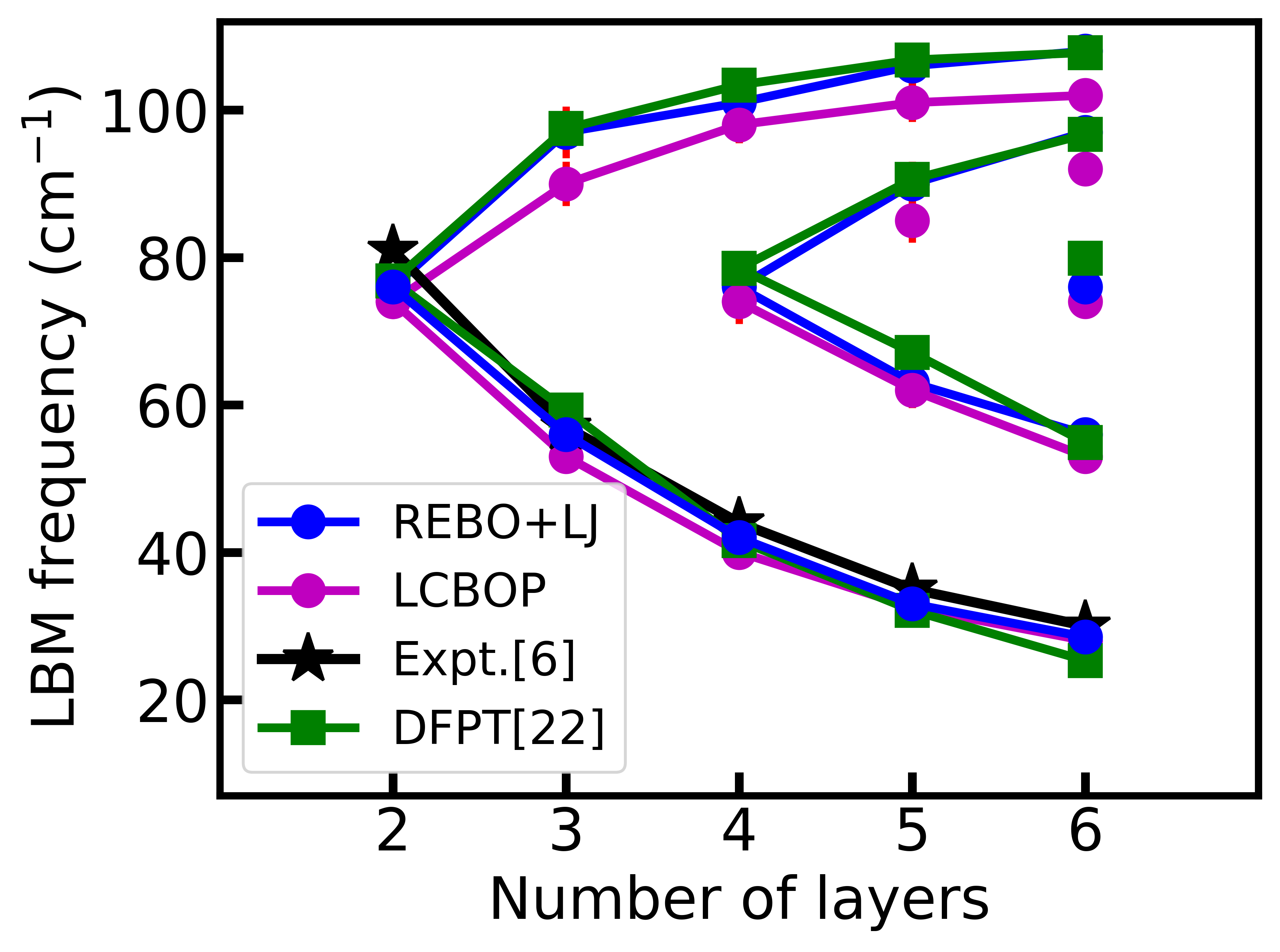}}
    \subfigure[]{\includegraphics[scale=0.38]{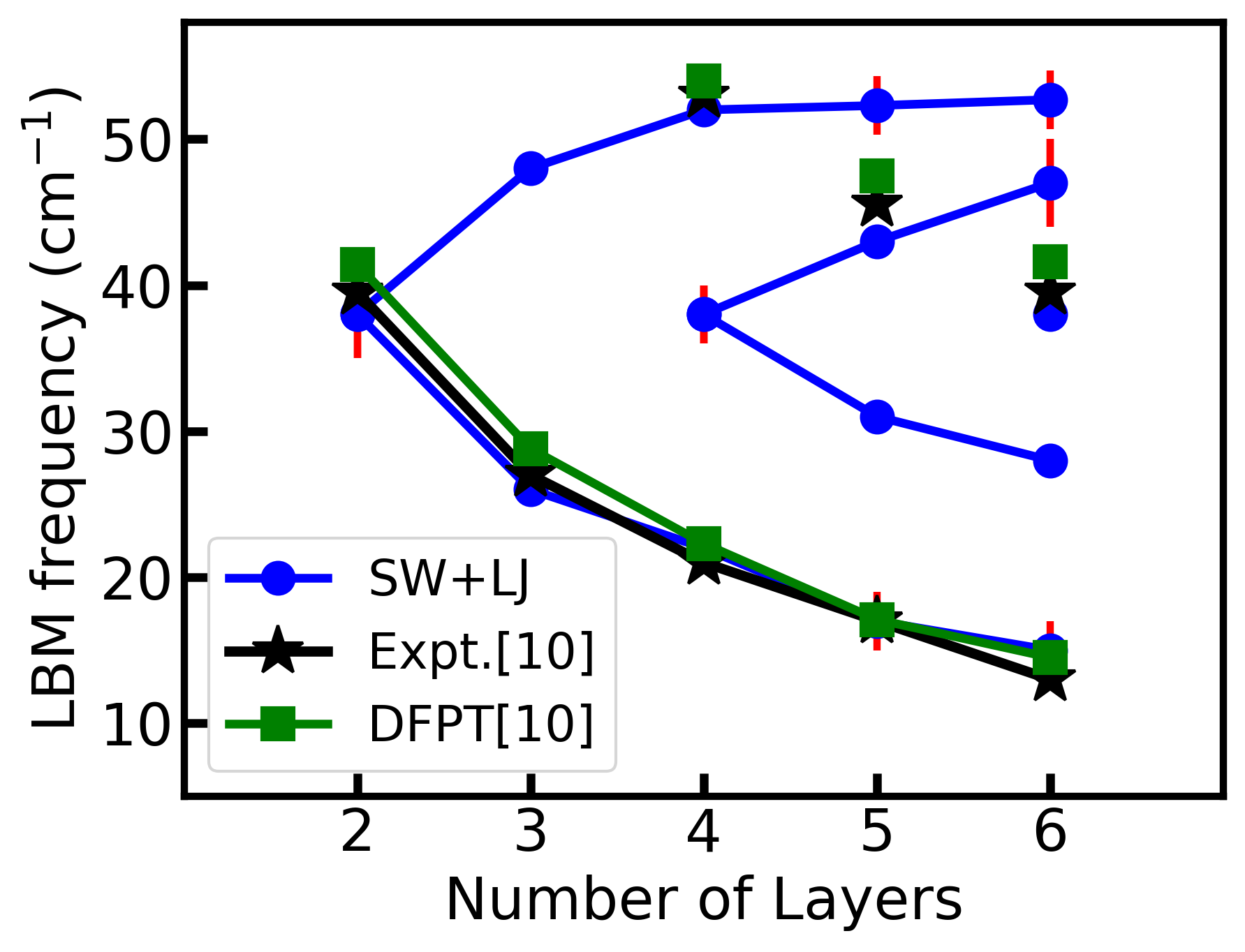}}
    \subfigure[]{\includegraphics[scale=0.38]{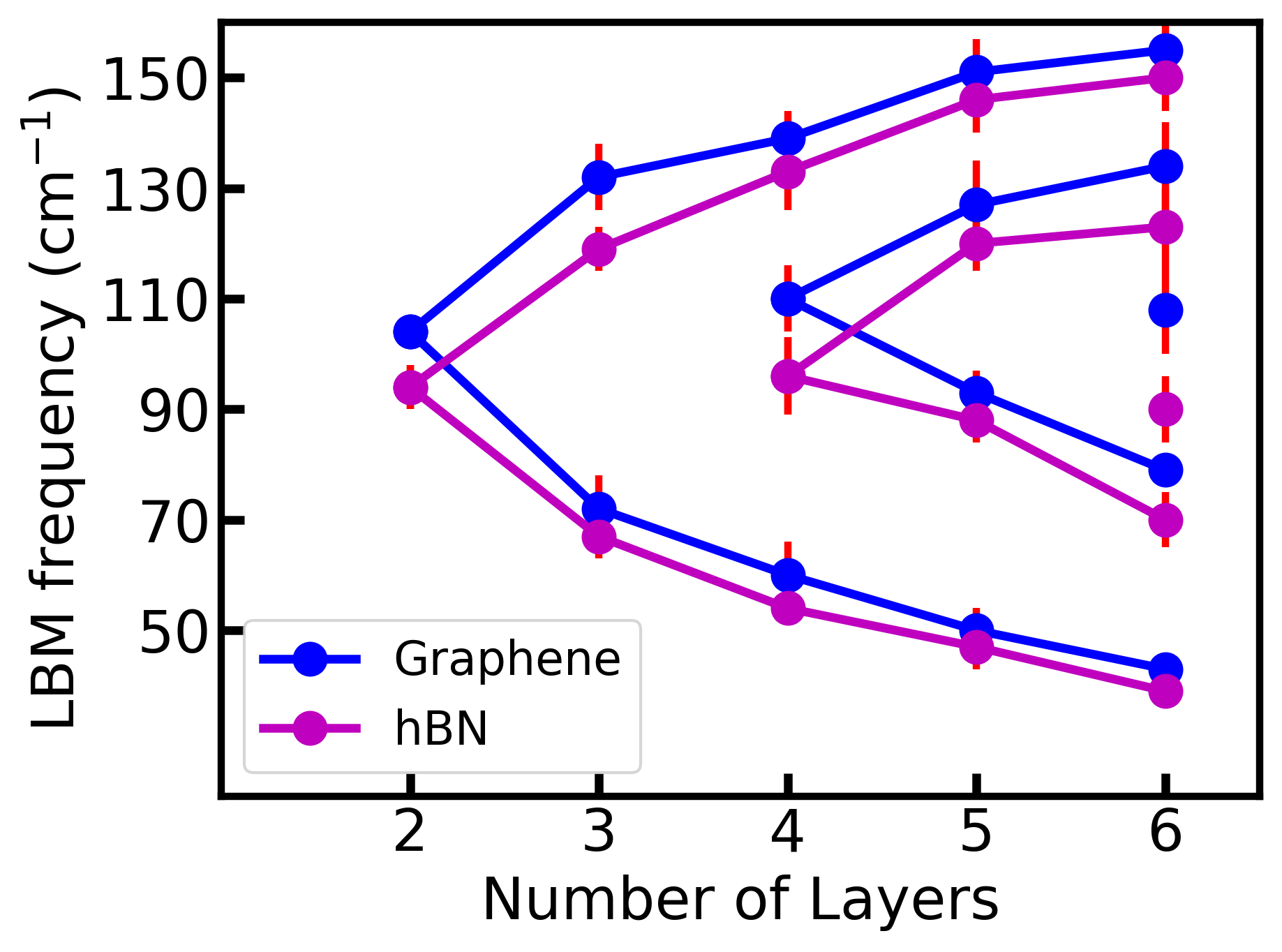}}
\caption{Thickness dependence of LBM frequencies.
(a) FL graphene 
(b) FL MoS$_2$ 
(c) Comparison of FL graphene and FL hBN calculated using
Dreiding \cite{mayo} forcefield. Solid lines are used as a guide to the eye. Vertical lines
(red) denote error bars.}
\end{figure*}    

The generalization of LBMs from BL to few layer (FL) system, can be done in
similar fashion as in Eqn.(1). Keeping only the nearest neighbor layer coupling
terms in FL system, we find the normalized eigenvectors and use them to compute all the 
height correlation functions \textit{explicitly} [See SI, section C, D]. We find, the \textit{mean}
mode of FL system behaves like a single layer, for the studied sample sizes.
Similar to the case of BL system, the \textit{fluctuation} modes are identified
with LBMs. In Fig.2 we show the layer dependence of LBM frequencies for
graphene, MoS$_{2}$ and hBN. For a $n$ layer system, there are $(n-1)$ distinct
LBM frequencies. As can be seen from the figures, our results for
graphene (Fig.2(a)) and MoS$_2$ (Fig.2(b)) capture the layer dependence
accurately. The figures also show LBM frequencies using DFPT
\cite{sahaphonons,zhaoMoS2}. Experimental data for graphene are shown only for the lowest
LBM frequency as they dominate the Raman response \cite{luitemperature}. The
LBM with lowest frequency display an extraordinarily simple structure, where
constituent layers expand and compress with respect to the mid-layer (odd n) or
mid-point (even n). Qualitatively, this mode results to least restoring force,
hence, lowest frequency (For schematics see SI, section D.1 ). With Dreiding,
the frequencies are overestimated by $\sim 28 \%$ (Fig.2(c)) for FL graphene.
Although overestimated, the general trend for the thickness dependence of LBM
frequencies is similar for hBN and graphene, consistent with another prediction
\cite{michel}. We can't compare the LBM frequencies for hBN with the experimental data, as LBMs have not been characterized for hBN yet. 

Two simple traits of the evolution of frequencies with thickness of 2D samples
must be pointed out : (i) Upon
increasing the number of layers, interlayer coupling between nearest neighbor layers
remain almost constant (within the error bar), consistent with earlier report
\cite{sahaphonons}. Thus, by computing $\sigma$ from BL system and using a
simple linear chain model the dramatic redshifts of lowest frequency of LBMs
with thickness can be captured, \textit{without calculating explicitly} the height
fluctuation modes for FL sample. (ii) The effect of next-nearest neighbor
interaction, is found to be negligible, for all the simulated samples (see SI,
section C.1). If the coupling is significant enough, this method can be easily
applied by adding more terms to the bending energy and reevaluating the height
fluctuation correlations with normalized eigenvectors.     

So far, in the bending energy cost (Eqn.1) for a BL system, both the interlayer
and intralayer interaction terms are assumed to be harmonic. As is well known,
at constant $P$, upon heating the material, the volume changes. This change in
volume can be explained via inclusion of anharmonic terms in the Hamiltonian.
Also the change of phonon frequency ($\omega$) with $T$ can only be obtained
from the anharmonicities of the potential energy. We calculate the change of
LBM frequency with $T$,  $\chi = \frac{d\omega}{dT}$, the first order
temperature coefficient, to discern the anharmonic effects in the interlayer
interaction. In this regard, we also compute the change of interlayer
separation with $T$, $\alpha_{\perp} = \frac{1}{c}\frac{dc}{dT}$. All the
reported values are estimated for BL system, with $T$ well below the melting
point. 

Fig.3 shows the temperature dependence of interlayer separation and LBM
frequency for BL graphene with REBO+LJ and Dreiding FF. Our results show that the equilibrium spacing between layers, $c$ increases
with $T$ ($\alpha_{\perp} > 0$). Moreover, increasing $T$ leads to the
softening of the \textit{effective spring constant}, $\sigma$ of the harmonic
oscillator. This results a redshift ($\chi < 0$) of the LBM frequency, which can also be substantiated from table II. All the
anharmonic effects are automatically included in our calculation of $\sigma$.
In principle, $\chi$ can be grouped into two parts : ``self energy" shift due
to direct anharmonic coupling of the phonon modes, $\chi_{V}$ and shift because
of the volume change of the material, $\chi_{T}$ \cite{lingP,frequency_dependence}. As all our
simulations are carried out at constant $P$, both contributions are included in
the estimated $\chi$. The second order temperature coefficient is found to be
irrelevant in the studied temperature range. In table II, we have shown
$\alpha_{\perp}$, for BL systems and compared it to the bulk values \footnote{ The value of $\alpha_{\perp}$ and $\chi$ for MoS$_2$ is calculated by changing $T$ from 150 K to 450 K only}. 
With standard Dreiding parameters $\alpha_{\perp}$ is always underestimated
compared to more accurate FFs. We find that 
$\alpha_{\perp}$ for BL (with accurate FFs) is larger than that of the bulk (experiments). 
It is interesting to note that there is a difference
in order of magnitude for $\alpha_{\perp}$ $(\sim 10^{-5}$ K$^{-1}$) and the
in-plane expansion coefficient, $\alpha_{\parallel}$ $(\sim 10^{-6}$ K$^{-1}$).
This is consistent with earlier observations
\cite{mounet,hBN_expansion}. However, the fact that $\alpha_{\perp}$ is greater than $\alpha_{\parallel}$ is not very surprising. This is due to the difference in strengths of interlayer and intralayer interactions of 2D materials. 

\begin{figure}[!htp]
    \centering
    \includegraphics[scale=0.45]{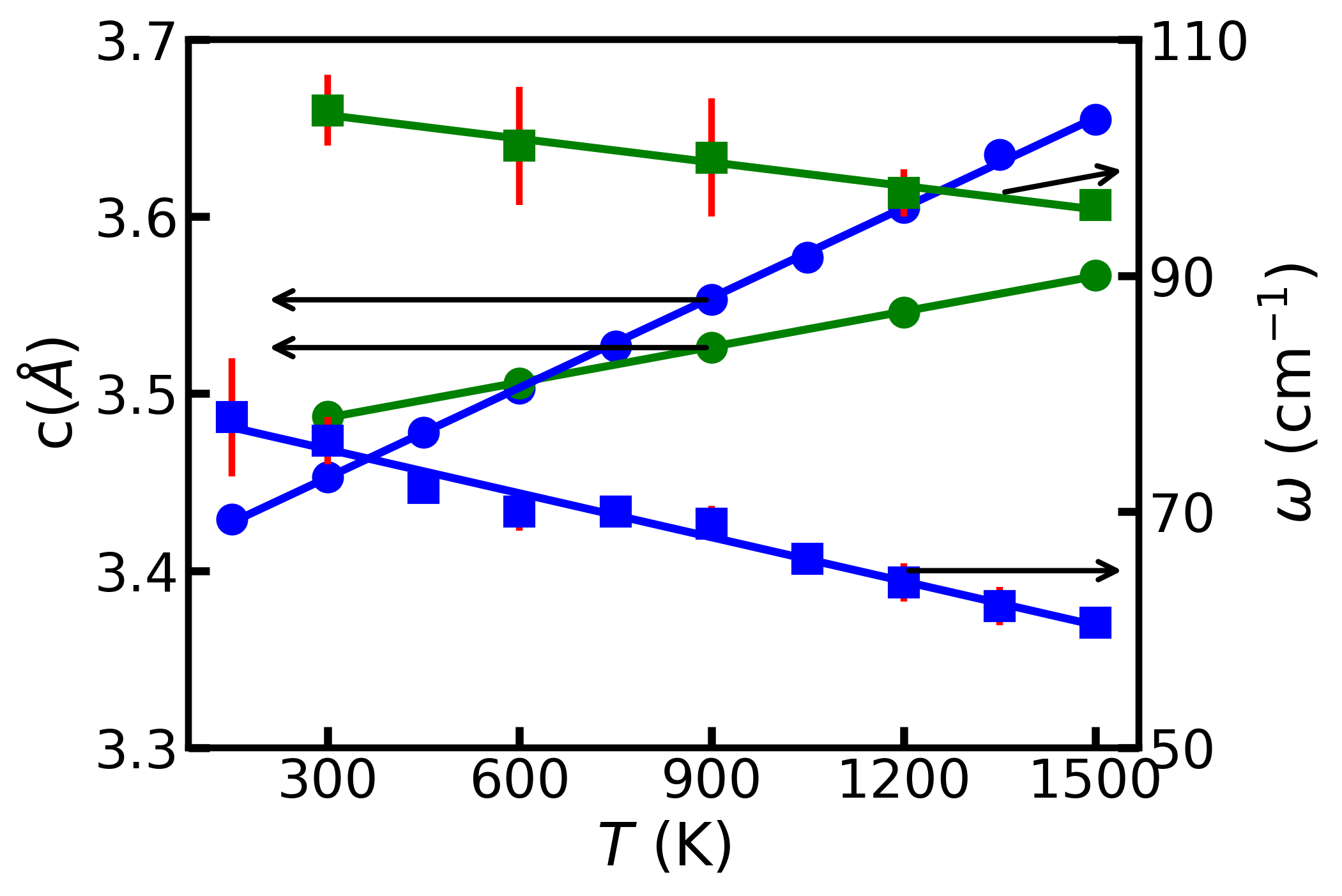}
    \caption{ Change of interlayer spacing, $c$ (left scale) and LBM frequency, $\omega$ (right scale) with $T$. Blue (green) circles present variation of $c$ using REBO+LJ (Dreiding) and blue squares (green squares) show evolution of $\omega$ using REBO+LJ (Dreiding). The solid lines represent linear fit to the data. }
\end{figure}
\begin{table}[htp]
    \caption{Effect of anharmonicity is to increase the interlayer spacing and redshift of LBM frequency. The thermal expansion coefficients are shown at $T=300$K.}
    \centering
    \begin{tabular}{|*{4}{c|}}
    \hline 
    \multicolumn{1}{|c|}{Material} & \multicolumn{1}{|c|}{Method} & \multicolumn{1}{|c|}{$\alpha_{\perp}$ (x10$^{-5}$ K$^{-1}$)} & \multicolumn{1}{|c|}{$\chi$ (x10$^{-3}$ cm$^{-1}$K${^{-1}}$)} \\ \hline
   BL  &  REBO+LJ & 4.9 $\pm$ 0.2 & -12.4 $\pm$ 0.8 \\ 
    Graphene & Dreiding & 1.9 $\pm$ 0.1  & -6 $\pm$ 1 \\ 
       & LCBOP & 6.2 $\pm$ 0.3  & -22.6 $\pm$ 0.9  \\ \hline 
       Graphite & Expt. \cite{graphite_expansion} & 2.7 & - \\ \hline 
     BL MoS$_2$ & SW+LJ & 2.4 $\pm$ 0.3 & -9.4 $\pm$ 1.4 \\ \hline
     Bulk MoS$_2$ & DFPT \cite{mos2_expansion} & 1.1 & - \\ \hline
     BL hBN & Dreiding  & 2.3 $\pm$ 0.1  & -6.7 $\pm$ 0.8  \\ \hline 
     Bulk hBN & Expt. \cite{hBN_expansion} & 3.77 & - \\ \hline 
    \end{tabular}
\end{table}

In summary, we have analyzed out of plane vibrations of 2D materials using a
combination of classical molecular dynamics simulations and membrane theory. We report our
results for three different classes of 2D materials, namely, graphene, MoS$_{2}$, hBN. We
provide, a consistent way to map LBMs of few layers of stacked 2D materials to
a simple linear chain model in the long-wavelength limit. The thickness
sensitivity of LBM frequencies at the $\Gamma$ point, are well captured and in
agreement with earlier reports. We also find, a redshift of LBM frequency upon
increasing $T$. We compute the interlayer separation thermal expansion
coefficient along with the shift in LBM frequency for BL systems. We show that
with accurate FFs LBM frequencies can be reliably estimated within this
simple picture. Our method also provides a framework to capture pressure or any other external environmental effects on LBM
frequencies.  This
study opens up the possibility of efficiently computing LBM frequencies
(including anharmonic effects) to characterize and understand properties of 2D
materials and their heterostructures.

The authors thank the Supercomputer Education and Research Center (SERC) at IISc for providing computational resources. The authors also thank Gaurav Kumar Gupta for comments on the manuscript.

\bibliography{mybib.bib}

\begin{thebibliography}{42}%
\makeatletter
\providecommand \@ifxundefined [1]{%
 \@ifx{#1\undefined}
}%
\providecommand \@ifnum [1]{%
 \ifnum #1\expandafter \@firstoftwo
 \else \expandafter \@secondoftwo
 \fi
}%
\providecommand \@ifx [1]{%
 \ifx #1\expandafter \@firstoftwo
 \else \expandafter \@secondoftwo
 \fi
}%
\providecommand \natexlab [1]{#1}%
\providecommand \enquote  [1]{``#1''}%
\providecommand \bibnamefont  [1]{#1}%
\providecommand \bibfnamefont [1]{#1}%
\providecommand \citenamefont [1]{#1}%
\providecommand \href@noop [0]{\@secondoftwo}%
\providecommand \href [0]{\begingroup \@sanitize@url \@href}%
\providecommand \@href[1]{\@@startlink{#1}\@@href}%
\providecommand \@@href[1]{\endgroup#1\@@endlink}%
\providecommand \@sanitize@url [0]{\catcode `\\12\catcode `\$12\catcode
  `\&12\catcode `\#12\catcode `\^12\catcode `\_12\catcode `\%12\relax}%
\providecommand \@@startlink[1]{}%
\providecommand \@@endlink[0]{}%
\providecommand \url  [0]{\begingroup\@sanitize@url \@url }%
\providecommand \@url [1]{\endgroup\@href {#1}{\urlprefix }}%
\providecommand \urlprefix  [0]{URL }%
\providecommand \Eprint [0]{\href }%
\providecommand \doibase [0]{http://dx.doi.org/}%
\providecommand \selectlanguage [0]{\@gobble}%
\providecommand \bibinfo  [0]{\@secondoftwo}%
\providecommand \bibfield  [0]{\@secondoftwo}%
\providecommand \translation [1]{[#1]}%
\providecommand \BibitemOpen [0]{}%
\providecommand \bibitemStop [0]{}%
\providecommand \bibitemNoStop [0]{.\EOS\space}%
\providecommand \EOS [0]{\spacefactor3000\relax}%
\providecommand \BibitemShut  [1]{\csname bibitem#1\endcsname}%
\let\auto@bib@innerbib\@empty
%</preamble>
\bibitem [{\citenamefont {Novoselov}\ \emph {et~al.}(2012)\citenamefont
  {Novoselov}, \citenamefont {Fal}, \citenamefont {Colombo}, \citenamefont
  {Gellert}, \citenamefont {Schwab}, \citenamefont {Kim} \emph
  {et~al.}}]{novoselovroadmap}%
  \BibitemOpen
  \bibfield  {author} {\bibinfo {author} {\bibfnamefont {K.~S.}\ \bibnamefont
  {Novoselov}}, \bibinfo {author} {\bibfnamefont {V.}~\bibnamefont {Fal}},
  \bibinfo {author} {\bibfnamefont {L.}~\bibnamefont {Colombo}}, \bibinfo
  {author} {\bibfnamefont {P.}~\bibnamefont {Gellert}}, \bibinfo {author}
  {\bibfnamefont {M.}~\bibnamefont {Schwab}}, \bibinfo {author} {\bibfnamefont
  {K.}~\bibnamefont {Kim}},  \emph {et~al.},\ }\href@noop {} {\bibfield
  {journal} {\bibinfo  {journal} {Nature}\ }\textbf {\bibinfo {volume} {490}},\
  \bibinfo {pages} {192} (\bibinfo {year} {2012})}\BibitemShut {NoStop}%
\bibitem [{\citenamefont {Butler}\ \emph {et~al.}(2013)\citenamefont {Butler},
  \citenamefont {Hollen}, \citenamefont {Cao}, \citenamefont {Cui},
  \citenamefont {Gupta}, \citenamefont {Guti{\'e}rrez}, \citenamefont {Heinz},
  \citenamefont {Hong}, \citenamefont {Huang}, \citenamefont {Ismach} \emph
  {et~al.}}]{beyondgraphene}%
  \BibitemOpen
  \bibfield  {author} {\bibinfo {author} {\bibfnamefont {S.~Z.}\ \bibnamefont
  {Butler}}, \bibinfo {author} {\bibfnamefont {S.~M.}\ \bibnamefont {Hollen}},
  \bibinfo {author} {\bibfnamefont {L.}~\bibnamefont {Cao}}, \bibinfo {author}
  {\bibfnamefont {Y.}~\bibnamefont {Cui}}, \bibinfo {author} {\bibfnamefont
  {J.~A.}\ \bibnamefont {Gupta}}, \bibinfo {author} {\bibfnamefont {H.~R.}\
  \bibnamefont {Guti{\'e}rrez}}, \bibinfo {author} {\bibfnamefont {T.~F.}\
  \bibnamefont {Heinz}}, \bibinfo {author} {\bibfnamefont {S.~S.}\ \bibnamefont
  {Hong}}, \bibinfo {author} {\bibfnamefont {J.}~\bibnamefont {Huang}},
  \bibinfo {author} {\bibfnamefont {A.~F.}\ \bibnamefont {Ismach}},  \emph
  {et~al.},\ }\href@noop {} {\bibfield  {journal} {\bibinfo  {journal} {ACS
  Nano}\ }\textbf {\bibinfo {volume} {7}},\ \bibinfo {pages} {2898} (\bibinfo
  {year} {2013})}\BibitemShut {NoStop}%
\bibitem [{\citenamefont {Geim}\ and\ \citenamefont
  {Grigorieva}(2013)}]{geimhetero}%
  \BibitemOpen
  \bibfield  {author} {\bibinfo {author} {\bibfnamefont {A.~K.}\ \bibnamefont
  {Geim}}\ and\ \bibinfo {author} {\bibfnamefont {I.~V.}\ \bibnamefont
  {Grigorieva}},\ }\href@noop {} {\bibfield  {journal} {\bibinfo  {journal}
  {Nature}\ }\textbf {\bibinfo {volume} {499}},\ \bibinfo {pages} {419}
  (\bibinfo {year} {2013})}\BibitemShut {NoStop}%
\bibitem [{\citenamefont {Tan}\ \emph {et~al.}(2012)\citenamefont {Tan},
  \citenamefont {Han}, \citenamefont {Zhao}, \citenamefont {Wu}, \citenamefont
  {Chang}, \citenamefont {Wang}, \citenamefont {Wang}, \citenamefont {Bonini},
  \citenamefont {Marzari}, \citenamefont {Pugno} \emph {et~al.}}]{tanshear}%
  \BibitemOpen
  \bibfield  {author} {\bibinfo {author} {\bibfnamefont {P.}~\bibnamefont
  {Tan}}, \bibinfo {author} {\bibfnamefont {W.}~\bibnamefont {Han}}, \bibinfo
  {author} {\bibfnamefont {W.}~\bibnamefont {Zhao}}, \bibinfo {author}
  {\bibfnamefont {Z.}~\bibnamefont {Wu}}, \bibinfo {author} {\bibfnamefont
  {K.}~\bibnamefont {Chang}}, \bibinfo {author} {\bibfnamefont
  {H.}~\bibnamefont {Wang}}, \bibinfo {author} {\bibfnamefont {Y.}~\bibnamefont
  {Wang}}, \bibinfo {author} {\bibfnamefont {N.}~\bibnamefont {Bonini}},
  \bibinfo {author} {\bibfnamefont {N.}~\bibnamefont {Marzari}}, \bibinfo
  {author} {\bibfnamefont {N.}~\bibnamefont {Pugno}},  \emph {et~al.},\
  }\href@noop {} {\bibfield  {journal} {\bibinfo  {journal} {Nature materials}\
  }\textbf {\bibinfo {volume} {11}},\ \bibinfo {pages} {294} (\bibinfo {year}
  {2012})}\BibitemShut {NoStop}%
\bibitem [{\citenamefont {Liang}\ \emph {et~al.}(2017)\citenamefont {Liang},
  \citenamefont {Zhang}, \citenamefont {Sumpter}, \citenamefont {Tan},
  \citenamefont {Tan},\ and\ \citenamefont {Meunier}}]{review_acsnano}%
  \BibitemOpen
  \bibfield  {author} {\bibinfo {author} {\bibfnamefont {L.}~\bibnamefont
  {Liang}}, \bibinfo {author} {\bibfnamefont {J.}~\bibnamefont {Zhang}},
  \bibinfo {author} {\bibfnamefont {B.~G.}\ \bibnamefont {Sumpter}}, \bibinfo
  {author} {\bibfnamefont {Q.-H.}\ \bibnamefont {Tan}}, \bibinfo {author}
  {\bibfnamefont {P.-H.}\ \bibnamefont {Tan}}, \ and\ \bibinfo {author}
  {\bibfnamefont {V.}~\bibnamefont {Meunier}},\ }\href@noop {} {\bibfield
  {journal} {\bibinfo  {journal} {ACS Nano}\ }\textbf {\bibinfo {volume}
  {11}},\ \bibinfo {pages} {11777} (\bibinfo {year} {2017})}\BibitemShut
  {NoStop}%
\bibitem [{\citenamefont {Lui}\ \emph {et~al.}(2014)\citenamefont {Lui},
  \citenamefont {Ye}, \citenamefont {Keiser}, \citenamefont {Xiao},\ and\
  \citenamefont {He}}]{luitemperature}%
  \BibitemOpen
  \bibfield  {author} {\bibinfo {author} {\bibfnamefont {C.~H.}\ \bibnamefont
  {Lui}}, \bibinfo {author} {\bibfnamefont {Z.}~\bibnamefont {Ye}}, \bibinfo
  {author} {\bibfnamefont {C.}~\bibnamefont {Keiser}}, \bibinfo {author}
  {\bibfnamefont {X.}~\bibnamefont {Xiao}}, \ and\ \bibinfo {author}
  {\bibfnamefont {R.}~\bibnamefont {He}},\ }\href@noop {} {\bibfield  {journal}
  {\bibinfo  {journal} {Nano Letters}\ }\textbf {\bibinfo {volume} {14}},\
  \bibinfo {pages} {4615} (\bibinfo {year} {2014})}\BibitemShut {NoStop}%
\bibitem [{\citenamefont {Lui}\ and\ \citenamefont {Heinz}(2013)}]{luiLBM1}%
  \BibitemOpen
  \bibfield  {author} {\bibinfo {author} {\bibfnamefont {C.~H.}\ \bibnamefont
  {Lui}}\ and\ \bibinfo {author} {\bibfnamefont {T.~F.}\ \bibnamefont
  {Heinz}},\ }\href@noop {} {\bibfield  {journal} {\bibinfo  {journal}
  {Physical Review B}\ }\textbf {\bibinfo {volume} {87}},\ \bibinfo {pages}
  {121404} (\bibinfo {year} {2013})}\BibitemShut {NoStop}%
\bibitem [{\citenamefont {He}\ \emph {et~al.}(2013)\citenamefont {He},
  \citenamefont {Chung}, \citenamefont {Delaney}, \citenamefont {Keiser},
  \citenamefont {Jauregui}, \citenamefont {Shand}, \citenamefont {Chancey},
  \citenamefont {Wang}, \citenamefont {Bao},\ and\ \citenamefont
  {Chen}}]{hegraphene}%
  \BibitemOpen
  \bibfield  {author} {\bibinfo {author} {\bibfnamefont {R.}~\bibnamefont
  {He}}, \bibinfo {author} {\bibfnamefont {T.-F.}\ \bibnamefont {Chung}},
  \bibinfo {author} {\bibfnamefont {C.}~\bibnamefont {Delaney}}, \bibinfo
  {author} {\bibfnamefont {C.}~\bibnamefont {Keiser}}, \bibinfo {author}
  {\bibfnamefont {L.~A.}\ \bibnamefont {Jauregui}}, \bibinfo {author}
  {\bibfnamefont {P.~M.}\ \bibnamefont {Shand}}, \bibinfo {author}
  {\bibfnamefont {C.}~\bibnamefont {Chancey}}, \bibinfo {author} {\bibfnamefont
  {Y.}~\bibnamefont {Wang}}, \bibinfo {author} {\bibfnamefont {J.}~\bibnamefont
  {Bao}}, \ and\ \bibinfo {author} {\bibfnamefont {Y.~P.}\ \bibnamefont
  {Chen}},\ }\href@noop {} {\bibfield  {journal} {\bibinfo  {journal} {Nano
  Letters}\ }\textbf {\bibinfo {volume} {13}},\ \bibinfo {pages} {3594}
  (\bibinfo {year} {2013})}\BibitemShut {NoStop}%
\bibitem [{\citenamefont {Zhang}\ \emph {et~al.}(2013)\citenamefont {Zhang},
  \citenamefont {Han}, \citenamefont {Wu}, \citenamefont {Milana},
  \citenamefont {Lu}, \citenamefont {Li}, \citenamefont {Ferrari},\ and\
  \citenamefont {Tan}}]{zhangMoS2}%
  \BibitemOpen
  \bibfield  {author} {\bibinfo {author} {\bibfnamefont {X.}~\bibnamefont
  {Zhang}}, \bibinfo {author} {\bibfnamefont {W.}~\bibnamefont {Han}}, \bibinfo
  {author} {\bibfnamefont {J.}~\bibnamefont {Wu}}, \bibinfo {author}
  {\bibfnamefont {S.}~\bibnamefont {Milana}}, \bibinfo {author} {\bibfnamefont
  {Y.}~\bibnamefont {Lu}}, \bibinfo {author} {\bibfnamefont {Q.}~\bibnamefont
  {Li}}, \bibinfo {author} {\bibfnamefont {A.}~\bibnamefont {Ferrari}}, \ and\
  \bibinfo {author} {\bibfnamefont {P.}~\bibnamefont {Tan}},\ }\href@noop {}
  {\bibfield  {journal} {\bibinfo  {journal} {Physical Review B}\ }\textbf
  {\bibinfo {volume} {87}},\ \bibinfo {pages} {115413} (\bibinfo {year}
  {2013})}\BibitemShut {NoStop}%
\bibitem [{\citenamefont {Zhao}\ \emph {et~al.}(2013)\citenamefont {Zhao},
  \citenamefont {Luo}, \citenamefont {Li}, \citenamefont {Zhang}, \citenamefont
  {Araujo}, \citenamefont {Gan}, \citenamefont {Wu}, \citenamefont {Zhang},
  \citenamefont {Quek}, \citenamefont {Dresselhaus} \emph {et~al.}}]{zhaoMoS2}%
  \BibitemOpen
  \bibfield  {author} {\bibinfo {author} {\bibfnamefont {Y.}~\bibnamefont
  {Zhao}}, \bibinfo {author} {\bibfnamefont {X.}~\bibnamefont {Luo}}, \bibinfo
  {author} {\bibfnamefont {H.}~\bibnamefont {Li}}, \bibinfo {author}
  {\bibfnamefont {J.}~\bibnamefont {Zhang}}, \bibinfo {author} {\bibfnamefont
  {P.~T.}\ \bibnamefont {Araujo}}, \bibinfo {author} {\bibfnamefont {C.~K.}\
  \bibnamefont {Gan}}, \bibinfo {author} {\bibfnamefont {J.}~\bibnamefont
  {Wu}}, \bibinfo {author} {\bibfnamefont {H.}~\bibnamefont {Zhang}}, \bibinfo
  {author} {\bibfnamefont {S.~Y.}\ \bibnamefont {Quek}}, \bibinfo {author}
  {\bibfnamefont {M.~S.}\ \bibnamefont {Dresselhaus}},  \emph {et~al.},\
  }\href@noop {} {\bibfield  {journal} {\bibinfo  {journal} {Nano Letters}\
  }\textbf {\bibinfo {volume} {13}},\ \bibinfo {pages} {1007} (\bibinfo {year}
  {2013})}\BibitemShut {NoStop}%
\bibitem [{\citenamefont {Boukhicha}\ \emph {et~al.}(2013)\citenamefont
  {Boukhicha}, \citenamefont {Calandra}, \citenamefont {Measson}, \citenamefont
  {Lancry},\ and\ \citenamefont {Shukla}}]{boukhichaMoS2}%
  \BibitemOpen
  \bibfield  {author} {\bibinfo {author} {\bibfnamefont {M.}~\bibnamefont
  {Boukhicha}}, \bibinfo {author} {\bibfnamefont {M.}~\bibnamefont {Calandra}},
  \bibinfo {author} {\bibfnamefont {M.-A.}\ \bibnamefont {Measson}}, \bibinfo
  {author} {\bibfnamefont {O.}~\bibnamefont {Lancry}}, \ and\ \bibinfo {author}
  {\bibfnamefont {A.}~\bibnamefont {Shukla}},\ }\href@noop {} {\bibfield
  {journal} {\bibinfo  {journal} {Physical Review B}\ }\textbf {\bibinfo
  {volume} {87}},\ \bibinfo {pages} {195316} (\bibinfo {year}
  {2013})}\BibitemShut {NoStop}%
\bibitem [{\citenamefont {Yan}\ \emph {et~al.}(2015)\citenamefont {Yan},
  \citenamefont {Xia}, \citenamefont {Wang}, \citenamefont {Liu}, \citenamefont
  {Kuo}, \citenamefont {Tay}, \citenamefont {Chen}, \citenamefont {Zhou},
  \citenamefont {Liu},\ and\ \citenamefont {Shen}}]{yanMoS2}%
  \BibitemOpen
  \bibfield  {author} {\bibinfo {author} {\bibfnamefont {J.}~\bibnamefont
  {Yan}}, \bibinfo {author} {\bibfnamefont {J.}~\bibnamefont {Xia}}, \bibinfo
  {author} {\bibfnamefont {X.}~\bibnamefont {Wang}}, \bibinfo {author}
  {\bibfnamefont {L.}~\bibnamefont {Liu}}, \bibinfo {author} {\bibfnamefont
  {J.-L.}\ \bibnamefont {Kuo}}, \bibinfo {author} {\bibfnamefont {B.~K.}\
  \bibnamefont {Tay}}, \bibinfo {author} {\bibfnamefont {S.}~\bibnamefont
  {Chen}}, \bibinfo {author} {\bibfnamefont {W.}~\bibnamefont {Zhou}}, \bibinfo
  {author} {\bibfnamefont {Z.}~\bibnamefont {Liu}}, \ and\ \bibinfo {author}
  {\bibfnamefont {Z.~X.}\ \bibnamefont {Shen}},\ }\href@noop {} {\bibfield
  {journal} {\bibinfo  {journal} {Nano Letters}\ }\textbf {\bibinfo {volume}
  {15}},\ \bibinfo {pages} {8155} (\bibinfo {year} {2015})}\BibitemShut
  {NoStop}%
\bibitem [{\citenamefont {Ling}\ \emph {et~al.}(2015)\citenamefont {Ling},
  \citenamefont {Liang}, \citenamefont {Huang}, \citenamefont {Puretzky},
  \citenamefont {Geohegan}, \citenamefont {Sumpter}, \citenamefont {Kong},
  \citenamefont {Meunier},\ and\ \citenamefont {Dresselhaus}}]{lingP}%
  \BibitemOpen
  \bibfield  {author} {\bibinfo {author} {\bibfnamefont {X.}~\bibnamefont
  {Ling}}, \bibinfo {author} {\bibfnamefont {L.}~\bibnamefont {Liang}},
  \bibinfo {author} {\bibfnamefont {S.}~\bibnamefont {Huang}}, \bibinfo
  {author} {\bibfnamefont {A.~A.}\ \bibnamefont {Puretzky}}, \bibinfo {author}
  {\bibfnamefont {D.~B.}\ \bibnamefont {Geohegan}}, \bibinfo {author}
  {\bibfnamefont {B.~G.}\ \bibnamefont {Sumpter}}, \bibinfo {author}
  {\bibfnamefont {J.}~\bibnamefont {Kong}}, \bibinfo {author} {\bibfnamefont
  {V.}~\bibnamefont {Meunier}}, \ and\ \bibinfo {author} {\bibfnamefont
  {M.~S.}\ \bibnamefont {Dresselhaus}},\ }\href@noop {} {\bibfield  {journal}
  {\bibinfo  {journal} {Nano Letters}\ }\textbf {\bibinfo {volume} {15}},\
  \bibinfo {pages} {4080} (\bibinfo {year} {2015})}\BibitemShut {NoStop}%
\bibitem [{\citenamefont {Zhao}\ \emph {et~al.}(2014)\citenamefont {Zhao},
  \citenamefont {Luo}, \citenamefont {Zhang}, \citenamefont {Wu}, \citenamefont
  {Bai}, \citenamefont {Wang}, \citenamefont {Jia}, \citenamefont {Peng},
  \citenamefont {Liu}, \citenamefont {Quek} \emph {et~al.}}]{zhaoBi2Te3}%
  \BibitemOpen
  \bibfield  {author} {\bibinfo {author} {\bibfnamefont {Y.}~\bibnamefont
  {Zhao}}, \bibinfo {author} {\bibfnamefont {X.}~\bibnamefont {Luo}}, \bibinfo
  {author} {\bibfnamefont {J.}~\bibnamefont {Zhang}}, \bibinfo {author}
  {\bibfnamefont {J.}~\bibnamefont {Wu}}, \bibinfo {author} {\bibfnamefont
  {X.}~\bibnamefont {Bai}}, \bibinfo {author} {\bibfnamefont {M.}~\bibnamefont
  {Wang}}, \bibinfo {author} {\bibfnamefont {J.}~\bibnamefont {Jia}}, \bibinfo
  {author} {\bibfnamefont {H.}~\bibnamefont {Peng}}, \bibinfo {author}
  {\bibfnamefont {Z.}~\bibnamefont {Liu}}, \bibinfo {author} {\bibfnamefont
  {S.~Y.}\ \bibnamefont {Quek}},  \emph {et~al.},\ }\href@noop {} {\bibfield
  {journal} {\bibinfo  {journal} {Physical Review B}\ }\textbf {\bibinfo
  {volume} {90}},\ \bibinfo {pages} {245428} (\bibinfo {year}
  {2014})}\BibitemShut {NoStop}%
\bibitem [{\citenamefont {Li}\ \emph {et~al.}(2017)\citenamefont {Li},
  \citenamefont {Wu}, \citenamefont {Ran}, \citenamefont {Lin}, \citenamefont
  {Liu}, \citenamefont {Zhao}, \citenamefont {Lu}, \citenamefont {Xiong},
  \citenamefont {Zhang}, \citenamefont {Huang}, \citenamefont {Zhang},\ and\
  \citenamefont {Tan}}]{mos2_graphene}%
  \BibitemOpen
  \bibfield  {author} {\bibinfo {author} {\bibfnamefont {H.}~\bibnamefont
  {Li}}, \bibinfo {author} {\bibfnamefont {J.-B.}\ \bibnamefont {Wu}}, \bibinfo
  {author} {\bibfnamefont {F.}~\bibnamefont {Ran}}, \bibinfo {author}
  {\bibfnamefont {M.-L.}\ \bibnamefont {Lin}}, \bibinfo {author} {\bibfnamefont
  {X.-L.}\ \bibnamefont {Liu}}, \bibinfo {author} {\bibfnamefont
  {Y.}~\bibnamefont {Zhao}}, \bibinfo {author} {\bibfnamefont {X.}~\bibnamefont
  {Lu}}, \bibinfo {author} {\bibfnamefont {Q.}~\bibnamefont {Xiong}}, \bibinfo
  {author} {\bibfnamefont {J.}~\bibnamefont {Zhang}}, \bibinfo {author}
  {\bibfnamefont {W.}~\bibnamefont {Huang}}, \bibinfo {author} {\bibfnamefont
  {H.}~\bibnamefont {Zhang}}, \ and\ \bibinfo {author} {\bibfnamefont {P.-H.}\
  \bibnamefont {Tan}},\ }\href@noop {} {\bibfield  {journal} {\bibinfo
  {journal} {ACS Nano}\ }\textbf {\bibinfo {volume} {11}},\ \bibinfo {pages}
  {11714} (\bibinfo {year} {2017})}\BibitemShut {NoStop}%
\bibitem [{\citenamefont {He}\ \emph {et~al.}(2016{\natexlab{a}})\citenamefont
  {He}, \citenamefont {van Baren}, \citenamefont {Yan}, \citenamefont {Xi},
  \citenamefont {Ye}, \citenamefont {Ye}, \citenamefont {Lu}, \citenamefont
  {Leong},\ and\ \citenamefont {Lui}}]{heNbSe2}%
  \BibitemOpen
  \bibfield  {author} {\bibinfo {author} {\bibfnamefont {R.}~\bibnamefont
  {He}}, \bibinfo {author} {\bibfnamefont {J.}~\bibnamefont {van Baren}},
  \bibinfo {author} {\bibfnamefont {J.-A.}\ \bibnamefont {Yan}}, \bibinfo
  {author} {\bibfnamefont {X.}~\bibnamefont {Xi}}, \bibinfo {author}
  {\bibfnamefont {Z.}~\bibnamefont {Ye}}, \bibinfo {author} {\bibfnamefont
  {G.}~\bibnamefont {Ye}}, \bibinfo {author} {\bibfnamefont {I.-H.}\
  \bibnamefont {Lu}}, \bibinfo {author} {\bibfnamefont {S.}~\bibnamefont
  {Leong}}, \ and\ \bibinfo {author} {\bibfnamefont {C.}~\bibnamefont {Lui}},\
  }\href@noop {} {\bibfield  {journal} {\bibinfo  {journal} {2D Materials}\
  }\textbf {\bibinfo {volume} {3}},\ \bibinfo {pages} {031008} (\bibinfo {year}
  {2016}{\natexlab{a}})}\BibitemShut {NoStop}%
\bibitem [{\citenamefont {He}\ \emph {et~al.}(2016{\natexlab{b}})\citenamefont
  {He}, \citenamefont {Yan}, \citenamefont {Yin}, \citenamefont {Ye},
  \citenamefont {Ye}, \citenamefont {Cheng}, \citenamefont {Li},\ and\
  \citenamefont {Lui}}]{heReS2}%
  \BibitemOpen
  \bibfield  {author} {\bibinfo {author} {\bibfnamefont {R.}~\bibnamefont
  {He}}, \bibinfo {author} {\bibfnamefont {J.-A.}\ \bibnamefont {Yan}},
  \bibinfo {author} {\bibfnamefont {Z.}~\bibnamefont {Yin}}, \bibinfo {author}
  {\bibfnamefont {Z.}~\bibnamefont {Ye}}, \bibinfo {author} {\bibfnamefont
  {G.}~\bibnamefont {Ye}}, \bibinfo {author} {\bibfnamefont {J.}~\bibnamefont
  {Cheng}}, \bibinfo {author} {\bibfnamefont {J.}~\bibnamefont {Li}}, \ and\
  \bibinfo {author} {\bibfnamefont {C.}~\bibnamefont {Lui}},\ }\href@noop {}
  {\bibfield  {journal} {\bibinfo  {journal} {Nano Letters}\ }\textbf {\bibinfo
  {volume} {16}},\ \bibinfo {pages} {1404} (\bibinfo {year}
  {2016}{\natexlab{b}})}\BibitemShut {NoStop}%
\bibitem [{\citenamefont {Lui}\ \emph {et~al.}(2015)\citenamefont {Lui},
  \citenamefont {Ye}, \citenamefont {Ji}, \citenamefont {Chiu}, \citenamefont
  {Chou}, \citenamefont {Andersen}, \citenamefont {Means-Shively},
  \citenamefont {Anderson}, \citenamefont {Wu}, \citenamefont {Kidd} \emph
  {et~al.}}]{luivdw}%
  \BibitemOpen
  \bibfield  {author} {\bibinfo {author} {\bibfnamefont {C.~H.}\ \bibnamefont
  {Lui}}, \bibinfo {author} {\bibfnamefont {Z.}~\bibnamefont {Ye}}, \bibinfo
  {author} {\bibfnamefont {C.}~\bibnamefont {Ji}}, \bibinfo {author}
  {\bibfnamefont {K.-C.}\ \bibnamefont {Chiu}}, \bibinfo {author}
  {\bibfnamefont {C.-T.}\ \bibnamefont {Chou}}, \bibinfo {author}
  {\bibfnamefont {T.~I.}\ \bibnamefont {Andersen}}, \bibinfo {author}
  {\bibfnamefont {C.}~\bibnamefont {Means-Shively}}, \bibinfo {author}
  {\bibfnamefont {H.}~\bibnamefont {Anderson}}, \bibinfo {author}
  {\bibfnamefont {J.-M.}\ \bibnamefont {Wu}}, \bibinfo {author} {\bibfnamefont
  {T.}~\bibnamefont {Kidd}},  \emph {et~al.},\ }\href@noop {} {\bibfield
  {journal} {\bibinfo  {journal} {Physical Review B}\ }\textbf {\bibinfo
  {volume} {91}},\ \bibinfo {pages} {165403} (\bibinfo {year}
  {2015})}\BibitemShut {NoStop}%
\bibitem [{\citenamefont {Perebeinos}\ \emph {et~al.}(2012)\citenamefont
  {Perebeinos}, \citenamefont {Tersoff},\ and\ \citenamefont
  {Avouris}}]{conductance}%
  \BibitemOpen
  \bibfield  {author} {\bibinfo {author} {\bibfnamefont {V.}~\bibnamefont
  {Perebeinos}}, \bibinfo {author} {\bibfnamefont {J.}~\bibnamefont {Tersoff}},
  \ and\ \bibinfo {author} {\bibfnamefont {P.}~\bibnamefont {Avouris}},\
  }\href@noop {} {\bibfield  {journal} {\bibinfo  {journal} {Physical Review
  Letters}\ }\textbf {\bibinfo {volume} {109}},\ \bibinfo {pages} {236604}
  (\bibinfo {year} {2012})}\BibitemShut {NoStop}%
\bibitem [{\citenamefont {Mahapatra}\ \emph {et~al.}(2017)\citenamefont
  {Mahapatra}, \citenamefont {Sarkar}, \citenamefont {Krishnamurthy},
  \citenamefont {Mukerjee},\ and\ \citenamefont {Ghosh}}]{phani_nano}%
  \BibitemOpen
  \bibfield  {author} {\bibinfo {author} {\bibfnamefont {P.~S.}\ \bibnamefont
  {Mahapatra}}, \bibinfo {author} {\bibfnamefont {K.}~\bibnamefont {Sarkar}},
  \bibinfo {author} {\bibfnamefont {H.~R.}\ \bibnamefont {Krishnamurthy}},
  \bibinfo {author} {\bibfnamefont {S.}~\bibnamefont {Mukerjee}}, \ and\
  \bibinfo {author} {\bibfnamefont {A.}~\bibnamefont {Ghosh}},\ }\href@noop {}
  {\bibfield  {journal} {\bibinfo  {journal} {Nano Letters}\ }\textbf {\bibinfo
  {volume} {17}},\ \bibinfo {pages} {6822} (\bibinfo {year}
  {2017})}\BibitemShut {NoStop}%
\bibitem [{Note1()}]{Note1}%
  \BibitemOpen
  \bibinfo {note} {There is no periodicity in the out of plane (z) direction of
  FL system. $\Gamma - A$ branch of the bulk counterpart is non existent in the
  Brillouin zone (BZ). One can show, the frequencies of LBMs at $\Gamma $ point
  for FL system, are associated with vibrations of their bulk correspondent
  along $\Gamma - A$ direction}\BibitemShut {NoStop}%
\bibitem [{\citenamefont {Saha}\ \emph {et~al.}(2008)\citenamefont {Saha},
  \citenamefont {Waghmare}, \citenamefont {Krishnamurthy},\ and\ \citenamefont
  {Sood}}]{sahaphonons}%
  \BibitemOpen
  \bibfield  {author} {\bibinfo {author} {\bibfnamefont {S.~K.}\ \bibnamefont
  {Saha}}, \bibinfo {author} {\bibfnamefont {U.}~\bibnamefont {Waghmare}},
  \bibinfo {author} {\bibfnamefont {H.}~\bibnamefont {Krishnamurthy}}, \ and\
  \bibinfo {author} {\bibfnamefont {A.}~\bibnamefont {Sood}},\ }\href@noop {}
  {\bibfield  {journal} {\bibinfo  {journal} {Physical Review B}\ }\textbf
  {\bibinfo {volume} {78}},\ \bibinfo {pages} {165421} (\bibinfo {year}
  {2008})}\BibitemShut {NoStop}%
\bibitem [{\citenamefont {Plimpton}(1995)}]{plimpton}%
  \BibitemOpen
  \bibfield  {author} {\bibinfo {author} {\bibfnamefont {S.}~\bibnamefont
  {Plimpton}},\ }\href@noop {} {\bibfield  {journal} {\bibinfo  {journal}
  {Journal of Computational Physics}\ }\textbf {\bibinfo {volume} {117}},\
  \bibinfo {pages} {1} (\bibinfo {year} {1995})}\BibitemShut {NoStop}%
\bibitem [{\citenamefont {Los}\ and\ \citenamefont {Fasolino}(2003)}]{los2003}%
  \BibitemOpen
  \bibfield  {author} {\bibinfo {author} {\bibfnamefont {J.}~\bibnamefont
  {Los}}\ and\ \bibinfo {author} {\bibfnamefont {A.}~\bibnamefont {Fasolino}},\
  }\href@noop {} {\bibfield  {journal} {\bibinfo  {journal} {Physical Review
  B}\ }\textbf {\bibinfo {volume} {68}},\ \bibinfo {pages} {024107} (\bibinfo
  {year} {2003})}\BibitemShut {NoStop}%
\bibitem [{\citenamefont {Brenner}\ \emph {et~al.}(2002)\citenamefont
  {Brenner}, \citenamefont {Shenderova}, \citenamefont {Harrison},
  \citenamefont {Stuart}, \citenamefont {Ni},\ and\ \citenamefont
  {Sinnott}}]{brennersecond}%
  \BibitemOpen
  \bibfield  {author} {\bibinfo {author} {\bibfnamefont {D.~W.}\ \bibnamefont
  {Brenner}}, \bibinfo {author} {\bibfnamefont {O.~A.}\ \bibnamefont
  {Shenderova}}, \bibinfo {author} {\bibfnamefont {J.~A.}\ \bibnamefont
  {Harrison}}, \bibinfo {author} {\bibfnamefont {S.~J.}\ \bibnamefont
  {Stuart}}, \bibinfo {author} {\bibfnamefont {B.}~\bibnamefont {Ni}}, \ and\
  \bibinfo {author} {\bibfnamefont {S.~B.}\ \bibnamefont {Sinnott}},\
  }\href@noop {} {\bibfield  {journal} {\bibinfo  {journal} {Journal of
  Physics: Condensed Matter}\ }\textbf {\bibinfo {volume} {14}},\ \bibinfo
  {pages} {783} (\bibinfo {year} {2002})}\BibitemShut {NoStop}%
\bibitem [{\citenamefont {Girifalco}\ \emph {et~al.}(2000)\citenamefont
  {Girifalco}, \citenamefont {Hodak},\ and\ \citenamefont {Lee}}]{graphene_lj}%
  \BibitemOpen
  \bibfield  {author} {\bibinfo {author} {\bibfnamefont {L.}~\bibnamefont
  {Girifalco}}, \bibinfo {author} {\bibfnamefont {M.}~\bibnamefont {Hodak}}, \
  and\ \bibinfo {author} {\bibfnamefont {R.~S.}\ \bibnamefont {Lee}},\
  }\href@noop {} {\bibfield  {journal} {\bibinfo  {journal} {Physical Review
  B}\ }\textbf {\bibinfo {volume} {62}},\ \bibinfo {pages} {13104} (\bibinfo
  {year} {2000})}\BibitemShut {NoStop}%
\bibitem [{\citenamefont {Mayo}\ \emph {et~al.}(1990)\citenamefont {Mayo},
  \citenamefont {Olafson},\ and\ \citenamefont {Goddard}}]{mayo}%
  \BibitemOpen
  \bibfield  {author} {\bibinfo {author} {\bibfnamefont {S.~L.}\ \bibnamefont
  {Mayo}}, \bibinfo {author} {\bibfnamefont {B.~D.}\ \bibnamefont {Olafson}}, \
  and\ \bibinfo {author} {\bibfnamefont {W.~A.}\ \bibnamefont {Goddard}},\
  }\href@noop {} {\bibfield  {journal} {\bibinfo  {journal} {Journal of
  Physical Chemistry}\ }\textbf {\bibinfo {volume} {94}},\ \bibinfo {pages}
  {8897} (\bibinfo {year} {1990})}\BibitemShut {NoStop}%
\bibitem [{\citenamefont {Jiang}\ \emph {et~al.}(2013)\citenamefont {Jiang},
  \citenamefont {Park},\ and\ \citenamefont {Rabczuk}}]{jiangsw}%
  \BibitemOpen
  \bibfield  {author} {\bibinfo {author} {\bibfnamefont {J.-W.}\ \bibnamefont
  {Jiang}}, \bibinfo {author} {\bibfnamefont {H.~S.}\ \bibnamefont {Park}}, \
  and\ \bibinfo {author} {\bibfnamefont {T.}~\bibnamefont {Rabczuk}},\
  }\href@noop {} {\bibfield  {journal} {\bibinfo  {journal} {Journal of Applied
  Physics}\ }\textbf {\bibinfo {volume} {114}},\ \bibinfo {pages} {064307}
  (\bibinfo {year} {2013})}\BibitemShut {NoStop}%
\bibitem [{\citenamefont {Liang}\ \emph {et~al.}(2009)\citenamefont {Liang},
  \citenamefont {Phillpot},\ and\ \citenamefont {Sinnott}}]{liang2009}%
  \BibitemOpen
  \bibfield  {author} {\bibinfo {author} {\bibfnamefont {T.}~\bibnamefont
  {Liang}}, \bibinfo {author} {\bibfnamefont {S.~R.}\ \bibnamefont {Phillpot}},
  \ and\ \bibinfo {author} {\bibfnamefont {S.~B.}\ \bibnamefont {Sinnott}},\
  }\href@noop {} {\bibfield  {journal} {\bibinfo  {journal} {Physical Review
  B}\ }\textbf {\bibinfo {volume} {79}},\ \bibinfo {pages} {245110} (\bibinfo
  {year} {2009})}\BibitemShut {NoStop}%
\bibitem [{\citenamefont {Liang}\ \emph {et~al.}(2012)\citenamefont {Liang},
  \citenamefont {Phillpot},\ and\ \citenamefont {Sinnott}}]{liang2012erratum}%
  \BibitemOpen
  \bibfield  {author} {\bibinfo {author} {\bibfnamefont {T.}~\bibnamefont
  {Liang}}, \bibinfo {author} {\bibfnamefont {S.~R.}\ \bibnamefont {Phillpot}},
  \ and\ \bibinfo {author} {\bibfnamefont {S.~B.}\ \bibnamefont {Sinnott}},\
  }\href@noop {} {\bibfield  {journal} {\bibinfo  {journal} {Physical Review
  B}\ }\textbf {\bibinfo {volume} {85}},\ \bibinfo {pages} {199903} (\bibinfo
  {year} {2012})}\BibitemShut {NoStop}%
\bibitem [{\citenamefont {Nelson}\ \emph {et~al.}(2004)\citenamefont {Nelson},
  \citenamefont {Piran},\ and\ \citenamefont {Weinberg}}]{nelsonbook}%
  \BibitemOpen
  \bibfield  {author} {\bibinfo {author} {\bibfnamefont {D.}~\bibnamefont
  {Nelson}}, \bibinfo {author} {\bibfnamefont {T.}~\bibnamefont {Piran}}, \
  and\ \bibinfo {author} {\bibfnamefont {S.}~\bibnamefont {Weinberg}},\
  }\href@noop {} {\emph {\bibinfo {title} {Statistical mechanics of membranes
  and surfaces}}}\ (\bibinfo  {publisher} {World Scientific},\ \bibinfo {year}
  {2004})\BibitemShut {NoStop}%
\bibitem [{\citenamefont {Fasolino}\ \emph {et~al.}(2007)\citenamefont
  {Fasolino}, \citenamefont {Los},\ and\ \citenamefont
  {Katsnelson}}]{fasolinoripples}%
  \BibitemOpen
  \bibfield  {author} {\bibinfo {author} {\bibfnamefont {A.}~\bibnamefont
  {Fasolino}}, \bibinfo {author} {\bibfnamefont {J.}~\bibnamefont {Los}}, \
  and\ \bibinfo {author} {\bibfnamefont {M.~I.}\ \bibnamefont {Katsnelson}},\
  }\href@noop {} {\bibfield  {journal} {\bibinfo  {journal} {Nature Materials}\
  }\textbf {\bibinfo {volume} {6}},\ \bibinfo {pages} {858} (\bibinfo {year}
  {2007})}\BibitemShut {NoStop}%
\bibitem [{\citenamefont {Amorim}\ \emph {et~al.}(2014)\citenamefont {Amorim},
  \citenamefont {Rold{\'a}n}, \citenamefont {Cappelluti}, \citenamefont
  {Fasolino}, \citenamefont {Guinea},\ and\ \citenamefont
  {Katsnelson}}]{amorim}%
  \BibitemOpen
  \bibfield  {author} {\bibinfo {author} {\bibfnamefont {B.}~\bibnamefont
  {Amorim}}, \bibinfo {author} {\bibfnamefont {R.}~\bibnamefont {Rold{\'a}n}},
  \bibinfo {author} {\bibfnamefont {E.}~\bibnamefont {Cappelluti}}, \bibinfo
  {author} {\bibfnamefont {A.}~\bibnamefont {Fasolino}}, \bibinfo {author}
  {\bibfnamefont {F.}~\bibnamefont {Guinea}}, \ and\ \bibinfo {author}
  {\bibfnamefont {M.}~\bibnamefont {Katsnelson}},\ }\href@noop {} {\bibfield
  {journal} {\bibinfo  {journal} {Physical Review B}\ }\textbf {\bibinfo
  {volume} {89}},\ \bibinfo {pages} {224307} (\bibinfo {year}
  {2014})}\BibitemShut {NoStop}%
\bibitem [{Note2()}]{Note2}%
  \BibitemOpen
  \bibinfo {note} {This is the height-height correlation in momentum space. For
  convenience, we write it as height correlation function.}\BibitemShut {Stop}%
\bibitem [{\citenamefont {Le~Doussal}\ and\ \citenamefont
  {Radzihovsky}(1992)}]{leself}%
  \BibitemOpen
  \bibfield  {author} {\bibinfo {author} {\bibfnamefont {P.}~\bibnamefont
  {Le~Doussal}}\ and\ \bibinfo {author} {\bibfnamefont {L.}~\bibnamefont
  {Radzihovsky}},\ }\href@noop {} {\bibfield  {journal} {\bibinfo  {journal}
  {Physical Review Letters}\ }\textbf {\bibinfo {volume} {69}},\ \bibinfo
  {pages} {1209} (\bibinfo {year} {1992})}\BibitemShut {NoStop}%
\bibitem [{\citenamefont {Michel}\ and\ \citenamefont
  {Verberck}(2012)}]{michel}%
  \BibitemOpen
  \bibfield  {author} {\bibinfo {author} {\bibfnamefont {K.}~\bibnamefont
  {Michel}}\ and\ \bibinfo {author} {\bibfnamefont {B.}~\bibnamefont
  {Verberck}},\ }\href@noop {} {\bibfield  {journal} {\bibinfo  {journal}
  {Physical Review B}\ }\textbf {\bibinfo {volume} {85}},\ \bibinfo {pages}
  {094303} (\bibinfo {year} {2012})}\BibitemShut {NoStop}%
\bibitem [{\citenamefont {Postmus}\ \emph {et~al.}(1968)\citenamefont
  {Postmus}, \citenamefont {Ferraro},\ and\ \citenamefont
  {Mitra}}]{frequency_dependence}%
  \BibitemOpen
  \bibfield  {author} {\bibinfo {author} {\bibfnamefont {C.}~\bibnamefont
  {Postmus}}, \bibinfo {author} {\bibfnamefont {J.~R.}\ \bibnamefont
  {Ferraro}}, \ and\ \bibinfo {author} {\bibfnamefont {S.~S.}\ \bibnamefont
  {Mitra}},\ }\href@noop {} {\bibfield  {journal} {\bibinfo  {journal} {Phys.
  Rev.}\ }\textbf {\bibinfo {volume} {174}},\ \bibinfo {pages} {983} (\bibinfo
  {year} {1968})}\BibitemShut {NoStop}%
\bibitem [{Note3()}]{Note3}%
  \BibitemOpen
  \bibinfo {note} {The value of $\alpha _{\perp }$ and $\chi $ for MoS$_2$ is
  calculated by changing $T$ from 150 K to 450 K only}\BibitemShut {NoStop}%
\bibitem [{\citenamefont {Mounet}\ and\ \citenamefont
  {Marzari}(2005)}]{mounet}%
  \BibitemOpen
  \bibfield  {author} {\bibinfo {author} {\bibfnamefont {N.}~\bibnamefont
  {Mounet}}\ and\ \bibinfo {author} {\bibfnamefont {N.}~\bibnamefont
  {Marzari}},\ }\href@noop {} {\bibfield  {journal} {\bibinfo  {journal}
  {Physical Review B}\ }\textbf {\bibinfo {volume} {71}},\ \bibinfo {pages}
  {205214} (\bibinfo {year} {2005})}\BibitemShut {NoStop}%
\bibitem [{\citenamefont {Paszkowicz}\ \emph {et~al.}(2002)\citenamefont
  {Paszkowicz}, \citenamefont {Pelka}, \citenamefont {Knapp}, \citenamefont
  {Szyszko},\ and\ \citenamefont {Podsiadlo}}]{hBN_expansion}%
  \BibitemOpen
  \bibfield  {author} {\bibinfo {author} {\bibfnamefont {W.}~\bibnamefont
  {Paszkowicz}}, \bibinfo {author} {\bibfnamefont {J.}~\bibnamefont {Pelka}},
  \bibinfo {author} {\bibfnamefont {M.}~\bibnamefont {Knapp}}, \bibinfo
  {author} {\bibfnamefont {T.}~\bibnamefont {Szyszko}}, \ and\ \bibinfo
  {author} {\bibfnamefont {S.}~\bibnamefont {Podsiadlo}},\ }\href@noop {}
  {\bibfield  {journal} {\bibinfo  {journal} {Applied Physics A}\ }\textbf
  {\bibinfo {volume} {75}},\ \bibinfo {pages} {431} (\bibinfo {year}
  {2002})}\BibitemShut {NoStop}%
\bibitem [{\citenamefont {Bailey}\ and\ \citenamefont
  {Yates}(1970)}]{graphite_expansion}%
  \BibitemOpen
  \bibfield  {author} {\bibinfo {author} {\bibfnamefont {A.}~\bibnamefont
  {Bailey}}\ and\ \bibinfo {author} {\bibfnamefont {B.}~\bibnamefont {Yates}},\
  }\href@noop {} {\bibfield  {journal} {\bibinfo  {journal} {Journal of Applied
  Physics}\ }\textbf {\bibinfo {volume} {41}},\ \bibinfo {pages} {5088}
  (\bibinfo {year} {1970})}\BibitemShut {NoStop}%
\bibitem [{\citenamefont {Gan}\ and\ \citenamefont
  {Liu}(2016)}]{mos2_expansion}%
  \BibitemOpen
  \bibfield  {author} {\bibinfo {author} {\bibfnamefont {C.~K.}\ \bibnamefont
  {Gan}}\ and\ \bibinfo {author} {\bibfnamefont {Y.~Y.~F.}\ \bibnamefont
  {Liu}},\ }\href@noop {} {\bibfield  {journal} {\bibinfo  {journal} {Physical
  Review B}\ }\textbf {\bibinfo {volume} {94}},\ \bibinfo {pages} {134303}
  (\bibinfo {year} {2016})}\BibitemShut {NoStop}%
\end{thebibliography}%
\end{document}